\documentclass[twocolumn,superscriptaddress,aps]{revtex4-1}
\usepackage{graphicx}
\usepackage{array}
\usepackage{amsmath}
\usepackage{amssymb}
\usepackage{latexsym}
\usepackage{amsfonts}
\usepackage{mathrsfs}
\usepackage{color}
\usepackage{hyperref}

\begin{document}

\title{Operational quasiprobabilities for qudits}

\author{Junghee Ryu}
\email{rjhui@hanyang.ac.kr}
\affiliation{Institute of Theoretical Physics and Astrophysics, University of Gda\'{n}sk, 80-952 Gda\'{n}sk, Poland}
\affiliation{Department of Physics, Hanyang University, Seoul 133-791, Korea}
\affiliation{Center for Macroscopic Quantum Control, Seoul National University, Seoul, 151-742, Korea}
\affiliation{Research Institute for Natural Sciences, Hanyang University, Seoul 133-791, Korea}

\author{James Lim}
\affiliation{Department of Physics, Hanyang University, Seoul 133-791, Korea}
\affiliation{Center for Macroscopic Quantum Control, Seoul National University, Seoul, 151-742, Korea}
\affiliation{Research Institute for Natural Sciences, Hanyang University, Seoul 133-791, Korea}

\author{Sunghyuk Hong}
\affiliation{Department of Physics, Hanyang University, Seoul 133-791, Korea}

\author{Jinhyoung Lee}
\email{hyoung@hanyang.ac.kr}
\affiliation{Department of Physics, Hanyang University, Seoul 133-791, Korea}
\affiliation{Center for Macroscopic Quantum Control, Seoul National University, Seoul, 151-742, Korea}
\affiliation{School of Computational Sciences, Korea Institute for Advanced Study, Seoul 130-722, Korea}

\begin{abstract}
We propose an operational quasiprobability function for qudits, enabling a comparison between quantum and hidden-variable theories. We show that the quasiprobability function becomes positive semidefinite if consecutive measurement results are described by a hidden-variable model with locality and noninvasive measurability assumed. Otherwise, it is negative valued. The negativity depends on the observables to be measured as well as a given state, as the quasiprobability function is operationally defined. We also propose a marginal quasiprobability function and show that it plays the role of an entanglement witness for two qudits. In addition, we discuss an optical experiment of a polarization qubit to demonstrate its nonclassicality in terms of the quasiprobability function.
\end{abstract}
\maketitle

\newcommand{\bra}[1]{\left\langle #1\right|}
\newcommand{\ket}[1]{\left|#1\right\rangle}
\newcommand{\abs}[1]{\left|#1\right|}
\newcommand{\ave}[1]{\left<#1\right>}
\newcommand{\Tr}{\mbox{Tr}}


\section{Introduction}

Quantum physics exhibits striking features compared to classical physics such as complementarity, nonlocality, and entanglement. The most profound discoveries have been found in terms of Bell's inequality and the Leggett-Garg inequality, which local realistic and macrorealistic theories obey, respectively, but quantum theory can violate~\cite{Bell64,Clauser69,Leggett85}. A comparison of quantum and classical statistics has also provided significant insights into understanding quantum physics and separating its features from the classical. For instance, photons have been shown to exhibit antibunching effects that classical statistics of light cannot describe~\cite{Kimble77}. These quantum features are said to be nonclassical if the classical theory of light does not predict them. 

To compare quantum with classical statistics, the Wigner function has been employed to represent a joint distribution of position $x$ and momentum $p$ in phase space~\cite{Wigner32,Hillery84,Lee95}. Contrary to the classical statistics, it is not straightforward to define a joint probability distribution in quantum statistics due to the uncertainty relation between position and momentum; in quantum physics, when two observables are mutually complementary, one observable cannot be measured without disturbing the other. Due to the complementarity (or uncertainty) principle, the Wigner function is not always positive semidefinite and may be negative valued for some quantum states. As it is not allowed by any classical probability distribution, the negativity is regarded as a signature of the nonclassicality. The Wigner function, called a quasiprobability distribution function, has been generalized to discrete systems as quantum informatics has gained importance~\cite{Buot74,*Hannay80,*Wootters87,*Ferrie09,Gibbons04}. The generalized quasiprobability functions have been applied to the omnidirectional range of quantum information processing, including quantum tomography, quantum teleportation, and analysis of quantum algorithms \cite{Leonhardt95,*Leonhardt96a,*Leonhardt96b,Vaccaro90,*Miquel02a,*Miquel02b,*Paz02,*Koniorczyk01,*Durt08}.

The quasiprobability functions have made significant progress in their own context. Nevertheless, we need to be careful when directly comparing a quasiprobability function with its classical counterpart, as they can be associated with {\em different kinds of observations} even with the {\em same functional form}. For instance, consider a classical distribution function $P(x,p)$. A functional of $P(x,p)$,
\begin{equation}
	\int^{\infty}_{-\infty} \int^{\infty}_{-\infty} dx dp P(x,p) \, x p,
\end{equation}
is associated with the average value of the product $x p$ of position $x$ and momentum $p$ in a joint measurement. On the other hand, the same functional of Wigner function $W(x,p)$,
\begin{eqnarray}
	\int^{\infty}_{-\infty} \int^{\infty}_{-\infty} dx dp \, W(x,p)\, x p,
\end{eqnarray}
is associated with the quantum average of a Hermitian observable operator $\{\hat{x},\hat{p}\}=\frac{1}{2}(\hat{x} \hat{p}+\hat{p}\hat{x})$~\cite{Hillery84,Lee95}. This quantum average is not directly related to the average of $x p$ in the above joint measurement. It arises because the eigenvectors of the operator $\{\hat{x},\hat{p}\}$ are unequal to any joint (or consecutive) measurement of $\hat{x}$ and $\hat{p}$. Thus, Wigner function $W(x,p)$ and its classical counterpart $P(x,p)$ can be associated with different kinds of observations by the same functionals. We say that $W(x,p)$ is ``incommensurable'' with its classical counterpart $P(x,p)$~\footnote{We adopt the term ``incommensurate'' from Kuhn's incommensurability in T. S. Kuhn, {\it The Road Since Structure} (University of Chicago Press, Chicago, 2000).}. 
This incommensurability makes it difficult to interpret the nonclassicality of a quasiprobability distribution. This problem remains unsolved in the approaches of generalizing quasiprobability functions to discrete systems \cite{Galvao05,Cormick06}. On the other hand, consider a joint probability distribution in the sequence of measuring $p$ first and $x$ later~\cite{[{}] [{, Chap. 2.}] Peresbook}: 
\begin{eqnarray}
P_{\rm QM}(x,p)=P_{\rm QM}(x|p) P_{\rm QM}(p),
\end{eqnarray}
where $P_{\rm QM}(p)$ is a probability distribution of $p$, resulting from quantum theory, and $P_{\rm QM}(x|p)$ is a conditional probability of $x$ given $p$. Then, the functional of $\int dx dp \, P_{\rm QM}(x,p)\, x p$ is associated with the same observation as the classical counterpart, so that $P_{\rm QM}(x,p)$ is commensurate with the classical probability $P(x,p)$ in the consecutive measurements $p$ and $x$.

In this paper, we propose an operational approach to define a {\em commensurate} quasiprobability function, enabling a direct comparison between quantum and classical statistics. Here, the classical distribution is described by a local hidden-variable model with noninvasive measurability~\cite{Leggett85,Kofler13, *Zukowski12}. We show that for any classical distribution the commensurate quasiprobability function is positive semidefinite since it is a legitimate probability distribution. Based on the result, we classify classical and nonclassical states of a qubit by showing the negativity of the commensurate quasiprobability function. Remarkably, we find that the nonclassicality is {\em operationally} determined in the sense that the degree of the nonclassicality depends on the observables to be measured, e.g.,, a measurement setup, even for a given quantum state. In addition, we propose an optical experiment of a polarization qubit, where the nonclassicality of a photon can be revealed by using a commensurate quasiprobability function without any theoretical assumptions on photon loss and photon-detection inefficiency. Finally, we derive a sufficient condition for the entanglement of two qudits using a marginal quasiprobability function.

\section{Commensurate Quasiprobability Function} 

Suppose that $K$ possible (incompatible) observables $A_k$ are {\em selectively} and {\em consecutively} measured on a quantum system~\cite{[{}] [{, Chap. 2.}] Peresbook} (this is called sequential measurements~\cite{[{}] [{ and the references therein.}] Budroni13}).

\begin{figure*}[t]
	\centering
	\includegraphics[scale=0.8]{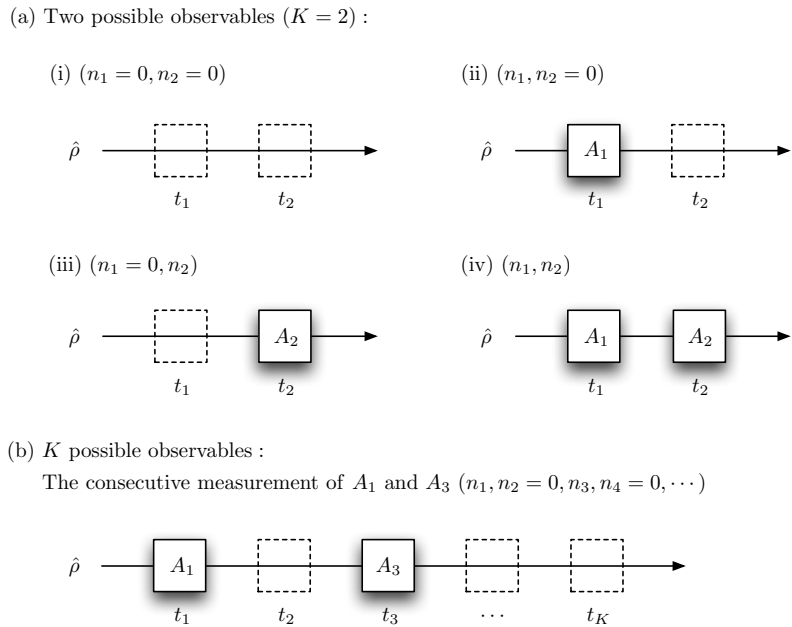}
	\caption{Schematic representation of measurement setups. When two observables $A_1$ and $A_2$ are selectively and consecutively measured, four measurement setups can be implemented as shown in (a): (i) no measurement, (ii) the measurement of $A_1$, (iii) the measurement of $A_2$, and (iv) the consecutive measurement of $A_1$ and $A_2$. When there are $K$ possible observables, $2^{K}$ measurement setups can be implemented. As an example, the consecutive measurement of $A_1$ and $A_3$ is displayed in (b).}
	\label{fig:schm}
\end{figure*}

Each nondegenerate measurement of an observable $A_k$ is performed at time $t_k$ with $t_1 < t_2 < \cdots < t_K$ if it is selected to be measured. In this case, depending on the selection of the observables, we implement one of $2^K$ measurement setups in which the selected observables are measured consecutively at different times; Fig.~\ref{fig:schm}(a) shows all possible measurement setups when there are only two observables $A_1$ and $A_2$. Each measurement setup is denoted by ${\bf n}=(n_1,n_2,\cdots,n_K)$, where $n_k\neq 0$ if the observable $A_k$ is selected to be measured and $n_k =0$ otherwise. We assume that each observable $A_k$ has $D$ possible outcomes denoted by $a_{k} \in \{0,1,\cdots,D-1\}$. Here, we consider both projective and positive operator-valued measure (POVM) measurements, implying that $D$ is independent of the dimension of the Hilbert space~\cite{Kraus83}.

In quantum theory, one needs to carefully describe the consecutive measurement of incompatible observables. For instance, suppose that two observables $A_{1}$ and $A_{3}$ are selected to be measured for a given quantum state $\hat{\rho}$, as shown in Fig.~\ref{fig:schm}(b). For simplicity, we assume that each observable $A_k$ is a projective measurement described by projectors $\Pi_{k} (a_k)$, which is associated with an outcome $a_k$ of measurement $A_k$. In this case, the measurement of $A_1$ can yield any outcome $a_1$ with the corresponding probability $p(a_1) = \mathrm{Tr}[\Pi_{1}(a_1)\hat{\rho}\,\Pi_1^{\dagger}(a_1)]$. Then measuring $A_3$ yields one of the outcomes $a_3$ with a conditional probability $p(a_3|a_1) = \Tr[\Pi_3 (a_3) \Pi_{1}(a_1) \hat{\rho}\, \Pi_{1}^{\dagger} (a_1) \Pi_{3}^{\dagger}(a_3) ]/p(a_1)$ depending on the outcome $a_1$ of the previous measurement $A_1$. This completes the description of the consecutive measurement of $A_{1}$ and $A_{3}$. Here, we employ a specific form of the expectation of the consecutive measurement $\chi(n_1,n_2=0,n_3, n_4=0,\cdots) =\sum_{a_1,a_3=0}^{D-1} \omega^{n_1 a_1 + n_3 a_3} \Tr[\Pi_3 (a_3) \Pi_{1}(a_1) \hat{\rho}\, \Pi_{1}^{\dagger} (a_1) \Pi_{3}^{\dagger}(a_3) ],$
with $\omega=e^{2\pi i/D}$ and $n_{k}\in\{1,2,\cdots,D-1\}$. Here, $\omega^{n_1 a_1 + n_3 a_3}$ is a possible value of the consecutive measurement corresponding to the probability $p(a_3|a_1) p(a_1)$ that the outcomes $a_1$ and $a_3$ will occur consecutively.

All expectations of such a form compose a function, which we call a characteristic function and denote by $\chi(\mathbf{n})$, with $\mathbf{n}=(n_1,n_2,\cdots,n_K)$. When the observables $A_k$ are not restricted to the projective measurements, we can employ POVM measurements. In this case, each measurement of $A_k$ is described by a set of the Kraus operators $\{\hat{A}_k(a_k)\}$ satisfying the positivity $\hat{A}_k^{\dagger} (a_k)\hat{A}_k (a_k)\ge 0$ and the completeness relation $\sum_{a_k=0}^{D-1} \hat{A}_k^\dag(a_k) \hat{A}_k(a_k) = \openone$, where $\openone$ denotes the identity operator. When an outcome $a_k\in\{0,1,\cdots,D-1\}$ occurs with a probability of $p(a_k)={\rm Tr}[\hat{A}_k(a_k)\hat{\rho}\hat{A}_{k}^{\dagger}(a_k)]$, the output state becomes $\hat{A}_k(a_k)\hat{\rho}\hat{A}_{k}^{\dagger}(a_k)/p(a_k)$. The characteristic function of the quantum state $\hat{\rho}$ is then given by
\begin{eqnarray}
	\chi(\mathbf{n}) = \mbox{Tr} \left[\mathcal{ T}\prod_{k=1}^K \left(\delta_{n_k,0} \mathcal{ I} + (1-\delta_{n_k,0})\sum_{a_k=0}^{D-1} \omega^{n_k a_k} \mathcal{ A}_k(a_k) \right)\hat{\rho}\right], \nonumber \\
	\label{eq:cfos}
\end{eqnarray}
with $\mathcal{ I}(\hat{\rho}) = \hat{\rho}$ and $\mathcal{ A}_k(a_k)(\hat{\rho}) = \hat{A}_k(a_k) \hat{\rho} \hat{A}_k^\dag(a_k)$. Here, $\delta_{n_k,0}$ represents the Kronecker delta defined by $\delta_{n_k,0}=1$ if $n_k=0$ and $\delta_{n_k,0}=0$ otherwise. The product of the superoperators is defined as their composition, e.g.,, $\mathcal{ A}_2 (a_2) \mathcal{ A}_1 (a_1) (\hat{\rho})= \hat{A}_2 (a_2) \hat{A}_1 (a_1) \hat{\rho} \hat{A}_1^\dag (a_1) \hat{A}_2^\dag (a_2)$, and $\mathcal{ T}$ denotes the chronological time-ordering operator defined by $\mathcal{ T} \mathcal{ A}_j \mathcal{ A}_k = \mathcal{ T} \mathcal{ A}_k \mathcal{ A}_j = \mathcal{ A}_k \mathcal{ A}_j$ if $t_k > t_j$, which describes the consecutive measurements of the observables $A_k$. As a simple case with two observables, $A_1$ and $A_2$, one can perform four measurement setups [see Fig.~\ref{fig:schm}(a)] and the characteristic function~\eqref{eq:cfos} of each setup is rewritten as
\begin{eqnarray}
\chi({0,0}) &=& \Tr[\hat{\rho}], \nonumber \\
\chi({n_1,0}) &=& \sum_{a_1 =0}^{D-1} \omega^{n_1 a_1} \Tr[\mathcal{A}_1 (a_1) \hat{\rho}], \nonumber \\
\chi({0,n_2}) &=& \sum_{a_2 =0}^{D-1} \omega^{n_2 a_2} \Tr[\mathcal{A}_2 (a_2) \hat{\rho}], \nonumber \\
\chi({n_1, n_2}) &=& \sum_{a_1,a_2=0}^{D-1} \omega^{n_1 a_1 + n_2 a_2} \Tr[\mathcal{A}_2 (a_2) \mathcal{A}_1 (a_1) \hat{\rho}]. \nonumber
\end{eqnarray}

We now propose a commensurate quasiprobability function defined by a discrete Fourier transformation of $\chi(\mathbf{n})$,
\begin{eqnarray}
	\mathcal{W}(\mathbf{a})
	\equiv\frac{1}{D^K} \sum_{\mathbf{n}=0}^{D-1} \omega^{- \mathbf{a} \cdot \mathbf{n}} \chi(\mathbf{n}),
	\label{eq:cqf}
\end{eqnarray}
where $\mathbf{a} = (a_1,a_2,\cdots,a_K)$, $\mathbf{a} \cdot \mathbf{n}=\sum_{k=1}^K a_k n_k$, with $a_k\in\{0,1,\cdots,D-1\}$, and $\sum_{\mathbf{n}=0}^{D-1}=\sum_{n_1=0}^{D-1}\sum_{n_2=0}^{D-1}\cdots\sum_{n_K=0}^{D-1}$. By definition, the characteristic function $\chi(\mathbf{n})$ is reproduced by the inverse Fourier transformation of $\mathcal{W}(\mathbf{a})$,
\begin{eqnarray}
	\chi(\mathbf{n})
	=\sum_{\mathbf{a}=0}^{D-1} \omega^{\mathbf{n} \cdot \mathbf{a}} \mathcal{W}(\mathbf{a}).
	\label{eq:idft}
\end{eqnarray}
It is notable that the functional of $\mathcal{W}(\mathbf{a})$ coincides with what it is supposed to represent, i.e.,, the expectation $\chi(\mathbf{n})$ of the consecutive measurement of incompatible observables (see Fig.~\ref{fig:schm}). Here $\omega^{\mathbf{n} \cdot \mathbf{a}}$ denotes a possible value of the measurement, while $\mathcal{W}(\mathbf{a})$ is placed at a position where the probability of measuring $\omega^{\mathbf{n} \cdot \mathbf{a}}$ would be located if the expectation $\chi(\mathbf{n})$ was described by a classical probability distribution. Depending on the quantum state $\hat{\rho}$ and observables $A_k$, a {\it nonnegative} quasiprobability function, i.e., $\mathcal{W}(\mathbf{a})\ge 0$, may not explain all the expectations $\chi(\mathbf{n})$. In Sec.~\ref{sec:nonclassicality}, we show that quantum theory allows the negativity of the quasiprobability function, i.e., $\mathcal{W}(\mathbf{a})<0$ for some ${\bf a}$, which is not allowed by any classical probability distribution. In addition, we find that the commensurate quasiprobability function is a real-valued function and satisfies the following conditions: (i) the sum of $\mathcal{W}(\mathbf{a})$ over all ${\bf a}$ is normalized, $\sum_{\mathbf{a}} \mathcal{W}(\mathbf{a}) = 1$, (ii) the sum of $\mathcal{W}(\mathbf{a})$ over a part of ${\bf a}$ gives the marginal quasiprobability of the rest, and (iii) the marginal quasiprobability of a single argument $a_k$ is equal to the probability of measuring $a_k$, $\mathcal{W}(a_k)={\rm Tr}[\hat{A}_{k}(a_k)\hat{\rho}\hat{A}_{k}^{\dagger}(a_k)]$. The second and third conditions play an important role in quantum tomography, as discussed in Refs.~\cite{Gibbons04,Leonhardt95}.

\section{Local Hidden-Variable Model with Noninvasive Measurability}\label{sec:local_model}

Various types of local hidden-variable models have been adopted in their own context, depending on their experimental circumstances~\cite{Bell64,Clauser69,Leggett85,Peres99,Kaszlikowski00,Leggett03,*Zukowski02,*Collins02,*Son06}. We take a local hidden-variable model with noninvasive measurability to compare the commensurate quasiprobability function with its classical counterpart. Local hidden-variable models have a common assumption that there exists a {\it nonnegative} probability distribution of the outcomes of all possible measurements. Our classical model additionally assumes the noninvasive measurability that it is possible, in principle, to determine the state of the system with an arbitrarily small perturbation on its subsequent dynamics~\cite{Kofler13, *Zukowski12}. This is understood as not only a spatially local but also {\it temporally} local hidden-variable model, which we call the classical model.

A classical expectation $\chi_\mathrm{cl}(\mathbf{n})$ is then given by
\begin{eqnarray}
	\chi_\mathrm{cl}(\mathbf{n}) &=&  \sum_{\mathbf{a}} \omega^{\mathbf{n} \cdot \mathbf{a}} p_\mathrm{cl}(\mathbf{a}),
	\label{eq:dhvtcfos}
\end{eqnarray}
where $p_\mathrm{cl}(\mathbf{a})$ is a classical joint probability of measuring outcomes ${\bf a}=(a_1,a_2,\cdots,a_K)$ when observables $A_k$ are selected to be measured. The Fourier transformation of the expectations $\chi_\mathrm{cl}(\mathbf{n})$ is then reduced to the classical probability distribution $p_\mathrm{cl}(\mathbf{a})\ge 0$. This implies that if the expectations of the consecutive measurements can be described by the classical model, the quasiprobability function $\mathcal{W}(\mathbf{a})$ is positive semidefinite, i.e., $\mathcal{W}(\mathbf{a})=p_\mathrm{cl}(\mathbf{a})\ge 0$, which is defined as the Fourier transformation of the expectations $\chi(\mathbf{n})$, as shown in Eq.~(\ref{eq:cqf}). As a contraposition, if the quasiprobability function $\mathcal{W}(\mathbf{a})$ is negative for some ${\bf a}$, the corresponding expectations $\chi(\mathbf{n})$ cannot be described by the classical model. In the next section, we show that quantum theory conflicts with the classical model and allows the negativity of the commensurate quasiprobability function.

\begin{figure*}[ht]
	\centering
	\includegraphics[scale=0.9]{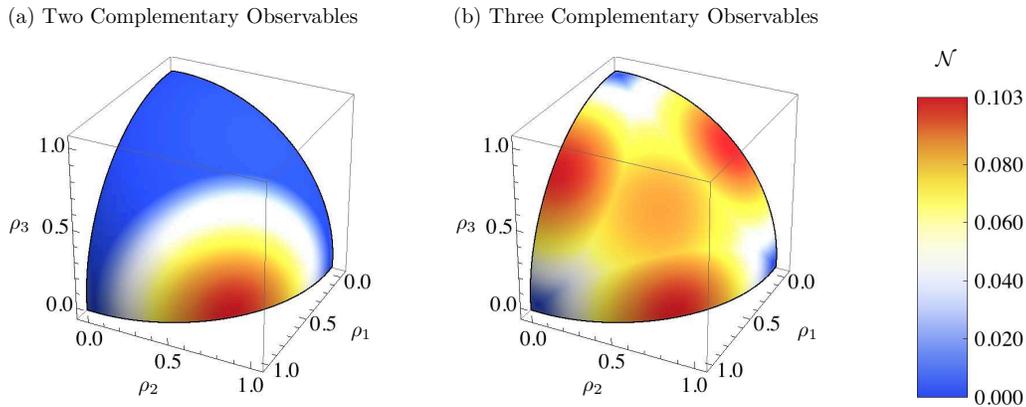}
	\caption{(Color online) Degree of nonclassicality $\mathcal{ N}$ of pure qubit states with mutually {\it unbiased} measurements. The Bloch vector $\vec{\rho}=(\rho_1,\rho_2,\rho_3)$ of a pure qubit state corresponds to a point on the Bloch sphere. In (a), where two complementary observables are employed, the nonclassicality is displayed as a function of the Bloch vector, i.e., $\mathcal{ N}(\rho_1,\rho_2,\rho_3)$. In (b), where three complementary observables are employed instead, the nonclassicality is displayed in a similar way. In both (a) and (b), the maximal nonclassicality is given by $\mathcal{ N}_{\rm max}=(\sqrt{2}-1)/4\approx 0.103$.}
	\label{fig:ncoftosq}
\end{figure*}
\section{Nonclassicality of a Qubit}\label{sec:nonclassicality}
\subsection{Complementary observables (mutually unbiased measurements)}

To demonstrate that quantum theory allows the negativity of the commensurate quasiprobability, we consider a $d$-dimensional quantum system called a qu$d$it with complementary observables $A_k$. The measurements of the complementary observables are described by mutually unbiased bases $\{| a_k \rangle\}$ satisfying the following conditions: (i) each projective measurement of $A_k$ has $d$ possible outcomes $(D=d)$, (ii) the orthonormal bases are mutually unbiased, i.e., $|\langle a_j | a_k \rangle|^2=1/d$ for all $j\neq k$ with $a_{j,k}\in\{0,1,\cdots,d-1\}$ (complementary relation)~\cite{Englert92,Lee03}, and (iii) the number of the complementary observables is no more than $d+1$ in a Hilbert space of dimension $d$ $(K\le d+1)$~\cite{Wootters89}. In this case, the commensurate quasiprobability function in Eq.~(\ref{eq:cqf}) is reduced to
\begin{eqnarray}
	\mathcal{W}(\mathbf{a})
	= \frac{1}{d^K} \left(1 + \sum_{k=1}^K  \vec{\alpha}_k(a_k) \cdot \vec{\rho}\right),
	\label{eq:gbvmub}
\end{eqnarray}
where $\vec{\rho}=(\rho_1, \rho_2, \cdots, \rho_{d^2 -1})$ represents the generalized Bloch vector of the quantum state $\hat{\rho}$ defined by $\rho_j=\Tr [\hat{\lambda}_j\hat{\rho}]$ with a complete orthogonal basis $\{\hat{\lambda}_j|j=0,1,2,\cdots,d^2-1\}$ satisfying $\hat{\lambda}_0=\openone$ and $\Tr [\hat{\lambda}_i\hat{\lambda}_j]=d\delta_{i,j}$. The generalized Bloch vectors of the complementary observables are defined similarly by $\vec{\alpha}_k (a_k)=\left(\alpha_k(a_k)_1,\alpha_k(a_k)_2,\cdots,\alpha_k(a_k)_{d^2-1}\right)$, with $\alpha_k (a_k)_j=\Tr [\hat{\lambda}_j | a_k \rangle\langle a_k|]$, and their inner products with $\vec{\rho}$ are defined by $\vec{\alpha}_k(a_k) \cdot \vec{\rho}=\sum_{j=1}^{d^2-1}\alpha_k(a_k)_j \rho_j$ (see the Appendix for details).

The quasiprobability $\mathcal{W}(\mathbf{a})$ in Eq.~(\ref{eq:gbvmub}) can be positive semidefinite for all ${\bf a}$ or can be negative for a part of ${\bf a}$. We quantify the degree of nonclassicality $\mathcal{ N}$ as a sum of the absolute values of the negative components of $\mathcal{W}(\mathbf{a})$:

\begin{eqnarray}
	\mathcal{N}
	=\frac{1}{2} \sum_{\mathbf{a}} \left[\left|\mathcal{W}(\mathbf{a})\right| - \mathcal{W}(\mathbf{a})\right].
\end{eqnarray}
Here, $\mathcal{ N}>0$ indicates the nonclassicality of the expectations $\chi(\mathbf{n})$. In this work, we call a quantum state classical if $\mathcal{ N}=0$ and nonclassical otherwise, which generally depends on the observables to be measured, as shown below.

To illustrate the nonclassical states, we consider a two-dimensional quantum system, known as a qubit $(d=2)$, with two complementary observables $(K=2)$. We employ the eigenbases of the Pauli spin operators $\hat{\sigma}_{x}$ and $\hat{\sigma}_{y}$ for modeling complementary observables such that $\hat{\sigma}_x| a_1 \rangle=(-1)^{a_1}| a_1 \rangle$ and $\hat{\sigma}_y| a_2 \rangle=(-1)^{a_2}| a_2 \rangle$ for $a_{1,2}\in\{0,1\}$. The Pauli spin operators are also used as a complete orthogonal basis, $\{\hat{\lambda}_0,\hat{\lambda}_1,\hat{\lambda}_2,\hat{\lambda}_3\}=\{\openone,\hat{\sigma}_x,\hat{\sigma}_y,\hat{\sigma}_z\}$. In this case, a quantum state is represented by, $\hat{\rho}=\frac{1}{2}(\openone+\vec{\rho}\cdot\vec{\lambda})$ with $\vec{\lambda}=(\hat{\lambda}_1,\hat{\lambda}_2,\hat{\lambda}_3)$ and $\vec{\rho}=(\rho_1,\rho_2,\rho_3)$ satisfying the normalization condition $\abs{\vec{\rho}\,}\le 1\Leftrightarrow\Tr[\hat{\rho}]=1$. Then the quasiprobability in Eq.~(\ref{eq:gbvmub}) is reduced to $\mathcal{W}(a_1,a_2) = [1 + (-1)^{a_1} \rho_1 + (-1)^{a_2} \rho_2]/4$. In Fig.~\ref{fig:ncoftosq}(a), the degree of nonclassicality $\mathcal{ N}$ for pure states is displayed as a function of the Bloch vector $\vec{\rho}$. Here the Bloch vectors of the pure states are characterized by $\abs{\vec{\rho}\,}=1$, which compose the Bloch sphere defined by $\rho_1^2+\rho_2^2+\rho_3^2=1$, where each pure state corresponds to a point on its surface. The maximal degree of nonclassicality over all possible qubit states is given by $\mathcal{ N}_\mathrm{max} = (\sqrt{2}-1)/4$, which is obtained by $\vec{\rho} = (\pm 1, \pm 1, 0)/\sqrt{2}$. For both pure and mixed states, $\mathcal{ N}>0$ if $\abs{\rho_1+\rho_2}>1$ or $\abs{\rho_1-\rho_2}>1$, and $\mathcal{ N}=0$ otherwise; in Fig.~\ref{fig:ncoftosq}(a), the regime of $\rho_1+\rho_2\le 1$ on the first octant of the Bloch sphere corresponds to the classical states of $\mathcal{ N}=0$, which are colored in blue (dark gray). On the other hand, when three complementary observables are employed $(K=3)$, the degree of nonclassicality of quantum states is dramatically changed, as shown in Fig.~\ref{fig:ncoftosq}(b). Here the Pauli spin operator $\hat{\sigma}_z$ is used for modeling an additional observable such that $\hat{\sigma}_z| a_3 \rangle=(-1)^{a_3}| a_3 \rangle$ for $a_{3}\in\{0,1\}$. In this case, the quasiprobability in Eq.~(\ref{eq:gbvmub}) is reduced to $\mathcal{W}(a_1,a_2,a_3)=[1 + (-1)^{a_1} \rho_1 + (-1)^{a_2} \rho_2+(-1)^{a_3} \rho_3]/8$, and all pure states become nonclassical except for $\vec{\rho}\in\{(\pm 1,0,0),(0,\pm 1,0),(0,0,\pm 1)\}$; these classical states coincide with those in Refs.~\cite{Galvao05,Cormick06}. The maximal degree of nonclassicality is $\mathcal{ N}_\mathrm{max} = (\sqrt{2}-1)/4$, as in the case of the two complementary observables, which is obtained by $\vec{\rho}\in\{(\pm 1,\pm 1,0)/\sqrt{2},(\pm 1,0,\pm 1)/\sqrt{2},(0,\pm 1,\pm 1)/\sqrt{2}\}$. It is notable that the nonclassicality of a given quantum state is determined {\it operationally} in our approach~\cite{Lee03,Brukner99,Bohr58} in the sense that a classical state in an experimental setup can be nonclassical in a different setup; in Fig.~\ref{fig:ncoftosq}, most classical states in Fig.~\ref{fig:ncoftosq}(a) become nonclassical in Fig.~\ref{fig:ncoftosq}(b) as the observables to be measured are changed from $\{\hat{\sigma}_x,\hat{\sigma}_y\}$ to $\{\hat{\sigma}_x,\hat{\sigma}_y,\hat{\sigma}_z\}$.

\subsection{Mutually biased measurements}

\begin{figure*}[ht]
	\centering
	\includegraphics[scale=0.9]{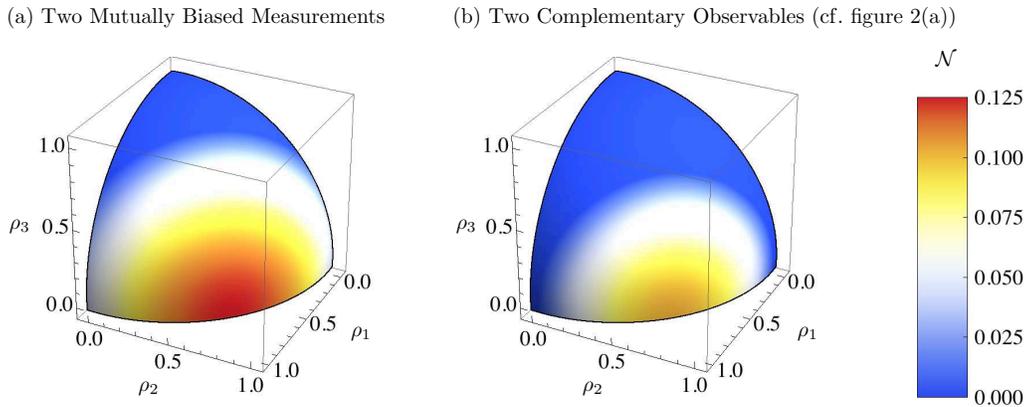}
	\caption{(Color online) Degree of nonclassicality $\mathcal{ N}$ of pure qubit states with mutually {\it biased} measurements. In (a), where two mutually {\it biased} measurements are employed, the nonclassicality is displayed as a function of the Bloch vector. In this case the maximal nonclassicality is given by $\mathcal{ N}_{\rm max}=0.125$, which is higher than the maximal value attainable by mutually {\it unbiased} measurements, $(\sqrt{2}-1)/4\approx 0.103$, as shown in (b).}
	\label{fig:notmubs}
\end{figure*}

So far we have considered complementary observables with mutually unbiased bases. However, the commensurate quasiprobability function proposed in this work can be applied to more general experimental setups with arbitrary projective and POVM measurements [see Eq.~(\ref{eq:cqf})]. As an example, we consider projective measurements with mutually {\it biased} bases and their influence on the nonclassicality of a qubit. Consider two projective measurements defined by $\hat{\sigma}_{1} = \cos (\theta) \hat{\sigma}_{x} - \sin (\theta) \hat{\sigma}_{y}$ and $\hat{\sigma}_{2} = -\sin (\theta) \hat{\sigma}_{x} + \cos (\theta) \hat{\sigma}_{y}$, where $\hat{\sigma}_{x}$ and $\hat{\sigma}_{y}$ are Pauli spin operators and $0<\theta<\pi/2$. In this case the observables $\hat{\sigma}_1$ and $\hat{\sigma}_2$ are not mutually complementary, i.e., $| \langle a_1 | a_2 \rangle |^2 \neq 1/2$ for all $a_{1,2}\in\{0,1\}$, where $\hat{\sigma}_1| a_1 \rangle=(-1)^{a_1}| a_1 \rangle$ and $\hat{\sigma}_2| a_2 \rangle=(-1)^{a_2}| a_2 \rangle$. The commensurate quasiprobability function in Eq.~{\rm(\ref{eq:cqf})} is then reduced to $\mathcal{W}(a_1,a_2) = [1 + (-1)^{a_1} \chi(1,0) + (-1)^{a_2} \chi(0,1)+(-1)^{a_1 +a_2}\chi(1,1)]/4$. We note that the quasiprobability function now contains the expectation of the consecutive measurement of $\hat{\sigma}_1$ and $\hat{\sigma}_2$, denoted by $\chi(1,1)$. This term does not vanish, in general, when the observables to be measured are biased, e.g., $| \langle a_1 | a_2 \rangle |^2 \neq 1/2$, contrary to the case of complementary observables. In Fig.~\ref{fig:notmubs}(a), the degree of nonclassicality $\mathcal{N}$ for pure qubit states is displayed for $\theta=\pi/12$. Here the maximal nonclassicality over all possible quantum states is given by $\mathcal{N}_{\mathrm{max}}=0.125$, which is obtained by $\vec{\rho}=(1,1,0)/\sqrt{2}$. It is notable that the mutually {\it biased} measurements considered here enhance the nonclassicality when compared to the case of complementary observables, where the nonclassicality is equal to or less than $(\sqrt{2}-1)/4\approx 0.103$. For comparison, we display the nonclassicality by complementary observables in Fig.~\ref{fig:notmubs}(b) [note that the color scale is different from that of Fig.~\ref{fig:ncoftosq}(a)].

\section{Commensurate Quasiprobability Function of a Polarization Qubit}
We propose an optical experiment to obtain a commensurate quasiprobability function of a polarization qubit that would measure the nonclassicality $\mathcal{ N}$ of a photon. To this end, we show that it is also a {\it positive semidefinite} probability distribution in a classical model, hidden-variables with noninvasive measurability. In our classical model, a photon is assumed to be a particle. By doing so, we exclude the quantum nature of antibunching for a photon, and we focus on other quantum characteristics that a photon may have but the classical model of hidden-variables cannot simulate.

\begin{figure}[t]
	\centering
	\includegraphics[scale=0.7]{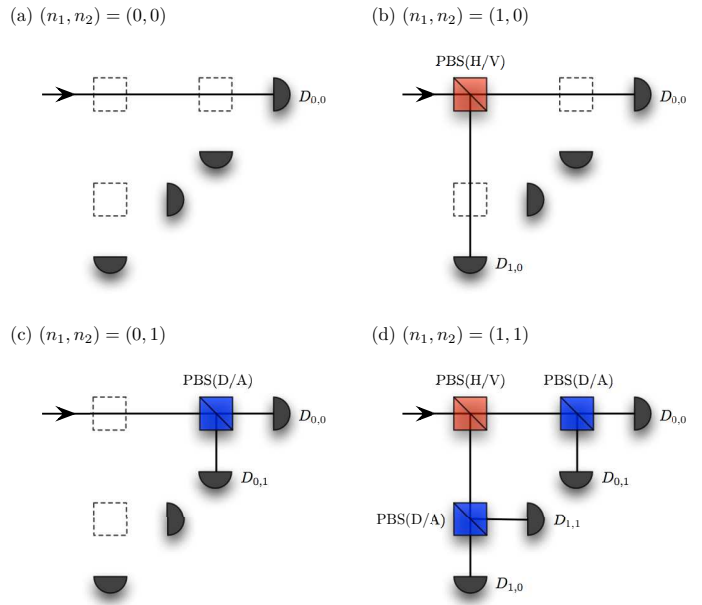}
	\caption{(Color online) An optical experiment of a polarization qubit. A red (light gray) square represents a polarizing beam splitter (PBS) that transmits a horizontally polarized photon (H) and reflects a vertically polarized one (V). A blue (dark gray) square denotes a PBS that transmits a diagonally polarized photon (D) and reflects an antidiagonally polarized one (A). A black half circle represents a photon detector and its position is denoted by $D_{a_1,a_2}$. The commensurate quasiprobability function can be obtained by Fourier transformation of the expectations of the measurement setups shown in (a)--(d).}
	\label{fig:sdepq}
\end{figure}

In Fig.~\ref{fig:sdepq}, we show all possible measurement setups for two possible observables, each of which measures horizontal and vertical polarizations (H and V) and diagonal and antidiagonal polarizations (D and A). In Figs.~\ref{fig:sdepq}(b) and \ref{fig:sdepq}(d), each red (light gray) square represents a polarizing beam splitter (PBS) that transmits a horizontally polarized photon and reflects a vertically polarized one. On the other hand, in Figs.~\ref{fig:sdepq}(c) and \ref{fig:sdepq}(d), each blue (dark gray) square denotes a PBS that transmits a diagonally polarized photon and reflects an antidiagonally polarized one. In Figs.~\ref{fig:sdepq}(a)--\ref{fig:sdepq}(d), black half circles represent photon detectors, each of which is placed at a position where a photon can be detected in the presence of beam splitters. The selection of the observables to be measured, which is denoted by $(n_1,n_2)$ with $n_{1,2}\in\{0,1\}$, determines the arrangement of beam splitters: $(n_1,n_2)=(0,0)$ corresponds to no polarization measurement [Fig.~\ref{fig:sdepq}(a)], $(n_1,n_2)=(1,0)$ corresponds to the measurement of H and V polarizations [Fig.~\ref{fig:sdepq}(b)], $(n_1,n_2)=(0,1)$ corresponds to the measurement of D and A polarizations [Fig.~\ref{fig:sdepq}(c)], and $(n_1,n_2)=(1,1)$ corresponds to the consecutive measurement of H and V and D and A polarizations [Fig.~\ref{fig:sdepq}(d)]. For each experimental setup, the associated expectation is given by
\begin{eqnarray}
	\tilde{\chi}_{\rm exp}(n_1,n_2)
	=\sum_{a_1, a_2=0}^{1}\omega^{n_1 a_1+n_2 a_2}f_{n_1,n_2}(a_1,a_2),
	\label{eq:characteristic_exp}
\end{eqnarray}
where $f_{n_1,n_2}(a_1,a_2)$ are the relative frequencies of photon counts at different detectors, i.e., $f_{n_1,n_2}(a_1,a_2)=N_{n_1,n_2}(a_1,a_2)/N_{n_1,n_2}^{\rm (det)}$, where $N_{n_1,n_2}(a_1,a_2)$ represents photon counts at a detector denoted by $D_{a_1,a_2}$ (see Fig.~\ref{fig:sdepq}) and $N_{n_1,n_2}^{\rm (det)}$ is the total number of detected events given by $N_{n_1,n_2}^{\rm (det)}=\sum_{a_1,a_2=0}^{1}N_{n_1,n_2}(a_1,a_2)$. The commensurate quasiprobability function is then given by the Fourier transformation of the expectations $\tilde{\chi}_{\rm exp}(n_1,n_2)$,
\begin{eqnarray}
	\tilde{\mathcal{ W}}_{\rm exp}(a_1,a_2)
	=\frac{1}{4}\sum_{n_1, n_2=0}^{1}\omega^{-a_1 n_1-a_2 n_2}\tilde{\chi}_{\rm exp}(n_1,n_2).
	\label{eq:quasiprobability_exp}
\end{eqnarray}
In the presence of photon loss and photon-detection inefficiency, some of the fired photons from the source will be lost or will not be detected in experiment. In this case the relative frequencies $f_{n_1,n_2}(a_1,a_2)$ in Eq.~(\ref{eq:characteristic_exp}) describe the {\it conditional} probabilities of detecting a photon, given that no photon loss takes place. It is notable that the conditional probabilities are the only quantities that can be determined {\it operationally} in experiment because the total number of photons fired from the source (or, equivalently, photon-loss probability) is generally unobservable.

We now show that the quasiprobability $\tilde{\mathcal{ W}}_{\rm exp}(a_1,a_2)$ consisting of the conditional probabilities is {\it positive semidefinite} if the expectations $\tilde{\chi}_{\rm exp}(n_1,n_2)$ can be described by a local hidden-variable model with noninvasive measurability. In the presence of photon loss and photon-detection inefficiency, the local hidden-variable model describes the expectations as follows:
\begin{eqnarray}
	\chi_{\rm cl}(n_1,n_2)
	=\sum_{a_1, a_2=0}^{1}\omega^{n_1 a_1+n_2 a_2}p_{\rm cl}(a_1,a_2,{\rm det}),
\end{eqnarray}
where $p_{\rm cl}(a_1,a_2,{\rm det})$ is the classical probability of detecting a photon at a detector denoted by $D_{a_1,a_2}$. The normalization condition of the probabilities is then given by
\begin{eqnarray}
	p_{\rm cl}^{({\rm loss})}+\sum_{a_1,a_2=0}^{1}p_{\rm cl}(a_1,a_2,{\rm det})=1,
\end{eqnarray}
where $p_{\rm cl}^{({\rm loss})}$ is the photon-loss probability. We divide the classical expectations by the total photon-detection probability defined by $p_{\rm cl}({\rm det})=\sum_{a_1,a_2=0}^{1}p_{\rm cl}(a_1,a_2,{\rm det})$, 
\begin{eqnarray}
	\tilde{\chi}_{\rm cl}(n_1,n_2)&=&\frac{\chi_{\rm cl}(n_1,n_2)}{p_{\rm cl}({\rm det})} \nonumber \\
	&=&\sum_{a_1,a_2=0}^{1}\omega^{n_1 a_1+n_2 a_2}p_{\rm cl}(a_1,a_2|{\rm det}),
\end{eqnarray}
where $p_{\rm cl}(a_1,a_2|{\rm det})=p_{\rm cl}(a_1,a_2,{\rm det})/p_{\rm cl}({\rm det})$ is the conditional probability of detecting a photon at a detector denoted by $D_{a_1,a_2}$, given that no photon loss takes place. The Fourier transformation of the classical expectations $\tilde{\chi}_{\rm cl}(n_1,n_2)$ is then reduced to the conditional probability distribution $p_{\rm cl}(a_1,a_2|{\rm det})$. This implies that when the expectations $\tilde{\chi}_{\rm exp}(n_1,n_2)$ measured in experiment can be described by the local hidden-variable model, the commensurate quasiprobability function is reduced to the classical distribution, i.e., $\tilde{\mathcal{ W}}_{\rm exp}(a_1,a_2)=p_{\rm cl}(a_1,a_2|{\rm det})\ge 0$, which is {\it positive semidefinite}. This implies that the negativity of the quasiprobability function indicates the nonclassicality of the photon, which can be determined {\it operationally} in experiment without any theoretical assumptions on photon loss. This is contrary to a common procedure in which the photon-loss probability is deduced or postulated from experimental data, which depends on the theoretical assumptions on photon loss. Our approach is similar in spirit to the Clauser-Horne inequality~\cite{Clauser74}.

\section{Marginal Quasiprobability Function as an Entanglement Witness}

When the commensurate quasiprobability function is applied to a composite system consisting of spatially separated subsystems, the quasiprobability function reveals two different types of nonclassicality. One is related to the temporal quantum correlation of each subsystem, which has been discussed in the previous sections. The other is the spatial quantum correlation between subsystems, such as entanglement~\cite{Bell64,Clauser69}. To distinguish the latter from the former, we now propose a marginal quasiprobability function of the composite system, which is positive semidefinite for all separable quantum states, implying that the negativity of the marginal quasiprobability function indicates the presence of entanglement.

We consider a composite system of two $d$-dimensional subsystems called qudits, each of which is distributed to one of the spatially separated observers Alice and Bob, respectively. Each qudit is selectively and consecutively measured by a complete set of complementary observables $(K=d+1)$, as is the case of the single system considered in Sec.~\ref{sec:nonclassicality}: here we assume that $d$ is a prime or a power of a prime number, for which the explicit forms of the mutually unbiased bases were developed by Wootters and Fields~\cite{Wootters89}. When Alice's and Bob's complementary observables are denoted by $A_k$ and $B_l$, respectively, with the associated orthonormal bases $\{| a_k \rangle\}$ and $\{| b_l \rangle\}$, the characteristic function of a composite quantum state $\hat{\rho}$ is given by
\begin{widetext}
\begin{eqnarray}
	\chi(\mathbf{m},\mathbf{n})
	=\Tr\left[\mathcal{ T}\prod_{k=1}^{d+1} \left( \delta_{m_{k},0} \mathcal{ I} + \left(1- \delta_{m_{k},0} \right) \sum_{a_k=0}^{d-1} \omega^{m_{k} a_k} \mathcal{ A}_{k}(a_k) \right)\otimes\prod_{l=1}^{d+1} \left( \delta_{n_{l},0} \mathcal{I}+ \left(1- \delta_{n_{l},0} \right) \sum_{b_l=0}^{d-1} \omega^{n_{l} b_l} \mathcal{ B}_{l}(b_l) \right) \hat{\rho}\right]\nonumber,
\end{eqnarray}
\end{widetext}
where $\mathcal{ A}_k(a_k)(\hat{\rho})=| a_k \rangle\langle a_k|\hat{\rho}| a_k \rangle\langle a_k|$ and $\mathcal{ B}_l(b_l)(\hat{\rho})=| b_l \rangle\langle b_l|\hat{\rho}| b_l \rangle\langle b_l|$. With the use of the complementary relation [see Eq.~\eqref{eq:complementarity}], the commensurate quasiprobability function defined by the Fourier transformation of $\chi(\mathbf{m},\mathbf{n})$ is given by
\begin{widetext}
\begin{eqnarray}
	\mathcal{W}(\mathbf{a},\mathbf{b})=\frac{1}{d^{2(d+1)}} \left(1 + \sum_{k=1}^{d+1} \vec{\alpha}_{k}(a_{k}) \cdot \vec{\rho}^{\,A}+\sum_{l=1}^{d+1} \vec{\beta}_{l}(b_{l})\cdot \vec{\rho}^{\,B}+\sum_{k,l=1}^{d+1}\vec{\alpha}_{k}(a_{k}) \cdot \mathbf{S} \cdot \vec{\beta}_{l}(b_{l}) \right)\label{eq:quasi_two_qudits},
\end{eqnarray}
\end{widetext}
where $\vec{\alpha}_{k}(a_{k})={\rm Tr}[(\vec{\lambda}^{A}\otimes\openone)(\ket{a_k}\bra{a_k}\otimes\openone)]$, $\vec{\rho}^{\,A}={\rm Tr}[(\vec{\lambda}^{A}\otimes\openone)\hat{\rho}]$, $\vec{\beta}_{l}(b_{l})={\rm Tr}[(\openone\otimes\vec{\lambda}^{B})(\openone\otimes\ket{b_l}\bra{b_l})]$, $\vec{\rho}^{\,B}={\rm Tr}[(\openone\otimes\vec{\lambda}^{B})\hat{\rho}]$, and $\mathbf{S}={\rm Tr}[(\vec{\lambda}^{A}\otimes\vec{\lambda}^{B})\hat{\rho}]$ in a generalized Bloch representation~\cite{Lee03}. Here $\vec{\rho}^{\,A}={\rm Tr}[(\vec{\lambda}^{A}\otimes\openone)\hat{\rho}]$ and $\mathbf{S}={\rm Tr}[(\vec{\lambda}^{A}\otimes\vec{\lambda}^{B})\hat{\rho}]$ stand for $\rho_{j}^{A}={\rm Tr}[(\hat{\lambda}_{j}^{A} \otimes \openone) \hat{\rho}]$ and $S_{jk}={\rm Tr}[(\hat{\lambda}_{j}^{A} \otimes \hat{\lambda}_{k}^{B})\hat{\rho}]$ for the sake of simplicity. The generalized Bloch vectors $\vec{\rho}^{\,A}$ and $\vec{\rho}^{\,B}$ describe the reduced quantum states of Alice's and Bob's qudits, respectively, while the generalized Bloch matrix $\mathbf{S}$ describes the spatial correlation between two qudits. Here the complete orthogonal basis for the Hilbert space of Alice's qudit, $\vec{\lambda}^A$, is generally different from that of Bob's qudit, $\vec{\lambda}^B$.

We now propose a marginal quasiprobability function:
\begin{eqnarray}
	\mathcal{W}_{m}(\mathbf{c})=\sum_{\bf a, b}\left(\prod_{j=1}^{d+1}\delta(c_j-a_{j}+b_{j}) \right) \, \mathcal{W}(\mathbf{a},\mathbf{b}),
	\label{eq:Marginal}
\end{eqnarray}
with $a_j,b_j,c_j\in\{0,1,\cdots,d-1\}$. Here $\delta(x)=1$ if $x\equiv 0 \mod d$ and $\delta(x)=0$ otherwise. When the complementary observables $A_k$ and $B_l$ are employed, by using $\mathcal{W}(\mathbf{a},\mathbf{b})$ in Eq.~(\ref{eq:quasi_two_qudits}) and $\sum_{a_k=0}^{d-1}\vec{\alpha}_{k}(a_k)=\vec{0}$~\cite{Lee03}, where $\vec{0}$ is a null vector, the marginal quasiprobability function is given by
\begin{widetext}
\begin{eqnarray}
	\mathcal{W}_{m}(\mathbf{c})
	&=&\frac{1}{d^{2(d+1)}}\sum_{\bf a}\left(1+\sum_{k,l=1}^{d+1}\vec{\alpha}_{k}(a_k)\cdot{\bf S}\cdot\vec{\beta}_{l}(a_l-c_l~{\rm mod}~d)\right) \nonumber \\
	&=&\frac{1}{d^{d+1}}\left(1+\frac{1}{d}\sum_{k=1}^{d+1}\sum_{x=0}^{d-1}\vec{\alpha}_{k}(x)\cdot{\bf S}\cdot\vec{\beta}_{k}(x-c_k~{\rm mod}~d)\right) =\frac{1}{d^{d+1}} \left(1+\Tr[{\bf SM(c)}]\right),
	\label{eq:marginal_ent}
\end{eqnarray}
\end{widetext}
where ${\bf M(c)}$ is a linear map from the generalized Bloch vector space to itself,
\begin{eqnarray}
	{\bf M(c)}=\frac{1}{d}\sum_{k=1}^{d+1}\sum_{x=0}^{d-1}\vec{\beta}_{k}(x-c_k~{\rm mod}~d)\vec{\alpha}_{k}(x).
\end{eqnarray}
For a separable state $\hat{\rho}^{\rm (sep)}=\sum_j p_j \hat{\rho}_{j}^{A}\otimes \hat{\rho}_{j}^{B}$, with $p_j\ge 0$ and $\sum_j p_j =1$, the marginal quasiprobability function in Eq.~(\ref{eq:marginal_ent}) is reduced to
\begin{eqnarray}
	\mathcal{W}_{m}^{\rm (sep)}(\mathbf{c}) = \frac{1}{d^{d+1}}\left(1+\sum_j p_j \vec{\rho}_{j}^{\,B} \cdot \mathbf{M}(\mathbf{c}) \cdot \vec{\rho}_{j}^{\,A} \right)\ge 0, \nonumber \\
\end{eqnarray}
where $\vec{\rho}_{j}^{\,A} = \Tr[\vec{\lambda}^{A} \hat{\rho}_{j}^{A}]$, $\vec{\rho}_{j}^{\,B} = \Tr[\vec{\lambda}^{B} \hat{\rho}_{j}^{B}]$, and $\vec{\rho}_{j}^{\,B} \cdot \mathbf{M}(\mathbf{c}) \cdot \vec{\rho}_{j}^{\,A} \ge -1$ for all $j$ as $\mathbf{M}(\mathbf{c}) \cdot \vec{\rho}_{j}^{\,A}$ is a generalized Bloch vector and the inner product between two generalized Bloch vectors is no less than $-1$~\cite{Lee03}. This implies that the marginal quasiprobability function is positive semidefinite for all separable quantum states. As a contraposition, this implies that a given quantum state is entangled if the associated marginal quasiprobability function $\mathcal{W}_{m}(\mathbf{c})$ is negative for some $\mathbf{c}$. As an example, we consider a Werner state,
\begin{eqnarray}
	\hat{\rho}^{\rm Werner}=p\ket{\rm \psi_{\rm MES}}\bra{\rm \psi_{\rm MES}}+(1-p)\frac{1}{d^2}\openone\otimes\openone,
\end{eqnarray}
where $0 \le p \le 1$ and $\ket{\rm \psi_{\rm MES}}=\frac{1}{\sqrt{d}}\sum_{n=0}^{d-1} |n\rangle \otimes |n\rangle$ is a maximally entangled state. Here we use specific forms of the complete orthogonal bases for Alice's and Bob's qudits, such that $\bra{n}\vec{\lambda}^{B}\ket{n'}=\bra{n'}\vec{\lambda}^{A}\ket{n}$ for all $n,n'\in\{0,1,\cdots,d-1\}$ in the Schmidt basis $\{\ket{n}\}$. The generalized Bloch matrix $\mathbf{S}$ then becomes the identity ${\bf I}$ multiplied by $p$, i.e., $S_{jk}=p\delta_{j,k}$ with the Kronecker delta $\delta_{j,k}=1$ if $j=k$ and $\delta_{j,k}=0$ otherwise,
\begin{widetext}
\begin{eqnarray}
	\mathbf{S}=p\bra{\psi_{\rm MES}}\vec{\lambda}^{A}\otimes\vec{\lambda}^{B}\ket{\psi_{\rm MES}}=\frac{p}{d}\sum_{n,n'=0}^{d-1}\bra{n}\vec{\lambda}^{A}\ket{n'}\bra{n}\vec{\lambda}^{B}\ket{n'}=\frac{p}{d}{\rm Tr}[\vec{\lambda}^{A}\vec{\lambda}^{A}]=p{\bf I},
\end{eqnarray}
\end{widetext}
leading to a simplified form of the marginal quasiprobability function,
\begin{widetext}
\begin{eqnarray}
	\mathcal{W}_{m}^{\rm Werner}(\mathbf{c})
	=\frac{1}{d^{d+1}} \left(1+p \frac{1}{d} \sum_{k=1} ^{d+1}\sum_{x=0}^{d-1} \vec{\alpha}_{k}(x) \cdot \vec{\beta}_{k}(x-c_k~{\rm mod}~d) \right)\label{eq:WERNER}
	\ge\frac{1}{d^{d+1}} \left[1-p(d+1) \right].
	\label{eq:INEQ}
\end{eqnarray}
\end{widetext}
Here the lower bound in Eq.~(\ref{eq:INEQ}) is due to the fact that the inner product between two generalized Bloch vectors is no less than -1~\cite{Lee03}, i.e., $\vec{\alpha}_{k}(x) \cdot \vec{\beta}_{k}(x-c_k~{\rm mod}~d)\ge -1$ for all $k$ and $x$. The equality holds when the eigenvectors of Alice's and Bob's complementary observables are given by $\{\ket{a_{k}}=\sum_{n=0}^{d-1}\phi_{kn}(a_k)\ket{n}\}$ and $\{\ket{b_{k}}=\sum_{n=0}^{d-1}\phi_{kn}^{*}(b_k)\ket{n}\}$ in the Schmidt basis $\{\ket{n}\}$. In this case, $\vec{\alpha}_k(x) \cdot \vec{\beta}_k (y)=d \delta_{x,y}-1$ due to the orthonormality condition $\vec{\alpha}_k(x) \cdot \vec{\alpha}_k (y)=d \delta_{x,y}-1$~\cite{Lee03} and $\vec{\alpha}_{k}(y)=\vec{\beta}_{k}(y)$:
\begin{widetext}
\begin{eqnarray}
	\vec{\alpha}_{k}(y)={\rm Tr}[(\vec{\lambda}^{A}\otimes\openone)(|y\rangle_k {}_k\langle y|\otimes\openone)] 
	={\rm Tr}[(\openone\otimes\vec{\lambda}^{B})(\openone\otimes|y\rangle_k {}_k\langle y|)]=\vec{\beta}_{k}(y).
\end{eqnarray}
\end{widetext}
This implies that when we set $c_k=1$ for all $k$ in Eq.~(\ref{eq:WERNER}), $\vec{\alpha}_{k}(x) \cdot \vec{\beta}_{k}(x-1~{\rm mod}~d)=-1$ for all $k$ and $x$, leading to the lower bound of the marginal quasiprobability function in Eq.~(\ref{eq:INEQ}). This shows that the marginal quasiprobability function becomes negative (for some $\mathbf{c}$) for the Werner states with $p>1/(d+1)$, implying that a Werner state is entangled if $p>1/(d+1)$. This sufficient condition for the presence of entanglement of the Werner states coincides with that in Ref.~\cite{Werner89}. These results indicate that the marginal quasiprobability function can be utilized as an entanglement witness~\cite{Terhal00}, where the negativity of the marginal quasiprobability is a sufficient condition for the presence of entanglement.

We call an observable $A$ an {\it entanglement witness} if $\Tr(A\, \hat{\rho}_{\rm s}) \geq 0$ for all separable states $\hat{\rho}_{\rm s}$ and $\Tr (A\, \hat{\rho}_{\rm e}) < 0$ for at least one entangled state $\hat{\rho}_{\rm e}$. Therefore, if we detect $\Tr(A\, \hat{\rho}_{\rm e}) < 0$, we know certainly that state $\hat{\rho}_{\rm e}$ is entangled. Entanglement witnesses are directly measurable quantities, so they are one of the most important methods for the analysis of entanglement in experiment. It is significant that our commensurate quasiprobability function naturally has such properties. By definition, every entanglement witnesses can detect some entangled state, but some witnesses are better than others for detecting entangled states. In this sense, we have an optimization problem of the entanglement witness. This is also the case for our commensurate quasiprobability function. We assumed mutually unbiased basis measurement so as to optimally detect entanglement for Werner states. Other types of entanglement require a different set of measurements to be optimal. Finding an optimal entanglement witness is a challenging problem in quantum information science.

\section{Summary}

We proposed a commensurate quasiprobability function for discrete systems, which is commensurate with its classical counterpart, enabling a direct comparison between quantum and classical statistics. We showed that the commensurate quasiprobability is positive semidefinite when the expectations of measurements can be described by a local hidden-variable model with noninvasive measurability. We demonstrated that quantum theory allows the negativity of the quasiprobability function and the negativity depends on both the quantum state and observables to be measured. In addition, we proposed an optical experiment of a polarization qubit and showed that the negativity of the quasiprobability function can be operationally determined in experiment without any theoretical assumptions on photon loss and photon-detection inefficiency. Finally, we proposed a marginal quasiprobability function for two qudits, which can be utilized as an entanglement witness. We showed that the marginal quasiprobability function is positive semidefinite for all separable quantum states and the negativity of the marginal quasiprobability function leads to a sufficient condition for the presence of entanglement of the Werner states. It would be interesting to apply a commensurate quasiprobability function to quantum information processing, for instance, to test if a given algorithm for quantum computation possesses nonclassical features or if it can be classically simulated by a classical (hidden-variable) model. It is an open question whether and/or how to define a commensurate quasiprobability function for a continuous-variable system, where its derivation might be difficult for its unbounded observables~\cite{Reed80}.

\acknowledgements
We are grateful to M. S. Kim, M. Tame, S.-W. Ji, J. Cho, C. Lee, W. Laskowski, M. Wie\'{s}niak, M. Paw\l owski, R. Rahaman, and M. \.{Z}ukowski for useful comments. This work was supported by the National Research Foundation of Korea (NRF) grant funded by the Korea government (MEST) (Grants No. 2010-0018295 and No. 2010-0015059). J.R. is supported by the Foundation for Polish Science TEAM project cofinanced by the EU European Regional Development Fund and a NCBiR-CHIST-ERA Project QUASAR.

\appendix

\section{}{\label{eq.1}}
We shall derive the form of commensurate quasiprobability function in Eq.~\eqref{eq:gbvmub} when local measurements are mutually unbiased bases. The quasiprobability function in Eq.~\eqref{eq:cqf} is rewritten as
\begin{eqnarray}
	\mathcal{W}(\mathbf{a}) = \Tr \left[ \mathcal{ T} \prod_{k=1}^K \left( \frac{1}{d}\,\mathcal{ I} + \Delta\mathcal{ A}_k(a_k) \right) [\hat{\rho}] \right],
	\label{eq:1.5}
\end{eqnarray}
where $\mathcal{ A}_k(a_k)(\hat{\rho})=| a_k \rangle \langle a_k|\hat{\rho}| a_k \rangle \langle a_k|$ and $\displaystyle\Delta\mathcal{ A}_k(a_k) = \mathcal{ A}_k(a_k) - \frac{1}{d}\sum_{a=0}^{d-1} \mathcal{ A}_k(a)$. The complementary relation between the observables is given by
\begin{eqnarray}
	\Tr \left[ \mathcal{ A}_k(a_k) \mathcal{ A}_j(a_j) \hat{\rho}\right] = \frac{1}{d}\Tr\left[\mathcal{ A}_j (a_j)\hat{\rho}\right], \nonumber \\
	\label{eq:complementarity}
\end{eqnarray}
for $t_k>t_j$. This leads to $\Tr\left[\mathcal{ T} \Delta \mathcal{ A}_k(a_k) \Delta\mathcal{ A}_j(a_j)\hat{\rho} \right]=0$, and it can be generalized to the case of an arbitrary combination of $\Delta\mathcal{ A}_k(a_k)$, i.e., $\Tr\left[\mathcal{ T} \Delta \mathcal{ A}_l(a_l) \cdots \Delta \mathcal{ A}_k(a_k) \cdots \Delta\mathcal{ A}_j(a_j)\hat{\rho} \right]=0$. This implies that only the zeroth and the first orders of $\Delta\mathcal{ A}_k(a_k)$ survive in Eq.~(\ref{eq:1.5}), while all higher orders vanish due to the complementary relation. The commensurate quasiprobability function is then simplified as
\begin{eqnarray}
	\mathcal{W}(\mathbf{a})
	&=& \frac{1}{d^{K}} \Tr \left[ \left( \mathcal{ I} + d \sum_{k=1}^K \Delta\mathcal{ A}_{k}(a_{k}) \right) \hat{\rho} \right] \nonumber \\
	&=& \frac{1}{d^K} \left(1 + \sum_{k=1}^K  \vec{\alpha}_k(a_k) \cdot \vec{\rho}\right),
\end{eqnarray}
with the generalized Bloch vectors $\vec{\alpha}_k (a_k)$ and $\vec{\rho}$.
\bibliography{reference}

\end{document}